\documentclass[twocolumn,floatfix,showpacs,prl,aps,tightenlines
]{revtex4}
\usepackage{graphicx}
\usepackage{color}
\usepackage{psfrag}
\usepackage{dcolumn}
\usepackage{bm}

\newcommand{\bea}{\begin{eqnarray}}
\newcommand{\eea}{\end{eqnarray}}
\newcommand{\beq}{\begin{equation}}
\newcommand{\eeq}{\end{equation}}
\newcommand{\KMS}{\rm km\,s^{-1}}

\begin{document}

\title{Extreme-Mass-Ratio-Black-Hole-Binary Evolutions \\
 with Numerical Relativity}

\author{Carlos O. Lousto}
\affiliation{Center for Computational Relativity and Gravitation,
School of Mathematical Sciences,
Rochester Institute of Technology, 85 Lomb Memorial Drive, Rochester,
 New York 14623, USA}

\author{Yosef Zlochower} 
\affiliation{Center for Computational Relativity and Gravitation,
School of Mathematical Sciences,
Rochester Institute of Technology, 85 Lomb Memorial Drive, Rochester,
 New York 14623, USA}

\date{\today}

\begin{abstract} 
We perform the first fully nonlinear numerical simulations of
black-hole binaries with mass ratios 100:1. Our technique is based on
the moving puncture formalism with a new gauge condition and an
optimal choice of the mesh refinement. The evolutions start with a
small nonspinning black hole just outside the ISCO that orbits twice
before plunging. We compute the gravitational radiation, as well as
the final remnant parameters, and find close agreement with
perturbative estimates. We briefly discuss the relevance of these
simulations for Advanced LIGO, third-generation ground-based
detectors, LISA observations, and self-force computations.
\end{abstract}

 \maketitle

\noindent{\it Introduction:}
The orbital evolution and computation of gravitational radiation from
black-hole binaries (BHB) in the small-mass-ratio limit remains one of the
most challenging problems in General Relativity. This was recognized
early on
by Regge and Wheeler over 50 years ago
\cite{Regge57}. Zerilli then completed the formulation of the first
order perturbations around a Schwarzschild BH in 1970
\cite{Zerilli:1971wd}. Three years later, Teukolsky
\cite{Teukolsky:1973ha} provided a new formalism to study
perturbations around Kerr BHs. In order to take into account
the decay of the orbit of the small BH  due to the emission of
gravitational radiation, second order effects have to be included in
those computations. This problem turned out to be very challenging, and
only since 1996 \cite{Mino:1996nk, Quinn:1996am} has there been  a consistent
formalism for the ``self-force'' corrections to the background geodesic
motion of a small BH orbiting around a larger one.  The explicit
implementation of such formalism into a computational scheme remains
challenging, although recent progress along this line is encouraging
\cite{Barack:2010tm}.

The dramatic breakthroughs in the numerical techniques to evolve
BHBs~\cite{Pretorius:2005gq, Campanelli:2005dd,
Baker:2005vv} transformed the field of Numerical Relativity (NR) and
we are now in a position to evolve binary systems in the intermediate
mass ratio regime. Two years ago the merger of spinning
\cite{Lousto:2008dn} binaries with mass ratio $q=m_1/m_2=1/8$ and
nonspinning binaries\cite{Gonzalez:2008bi} with $q=1/10$ were published.  More
recently, detailed long term evolutions of BHBs with $q=1/10$ and
$q=1/15$ were studied and validated against perturbation theory
\cite{Lousto:2010tb, Lousto:2010qx}.  In this Letter we present the
first fully nonlinear
numerical simulations of the merger of small-mass-ratio BHBs. As a
case study, we evolve a nonspinning BHB  with  mass ratio
$q=1/100$ for over two orbits prior to merger, and resolve the entire
waveform for three grid resolutions, proving numerical convergence of
the results.  The success of our approach is based on enhancements
of the moving puncture numerical techniques that  adapt the gauge and
grid structure to the small-mass-ratio limit.

The techniques described in this Letter can be used
in the spinning BHB case and for even smaller mass ratio inspirals.
This has important consequences for astrophysics and gravitational
wave observatories such as the second-generation Advanced LIGO
detector, third-generation
ground-based detectors, and LISA. Supermassive BH
collision at cosmological scales are most likely to occur in the mass
ratio range 1:10 - 1:100~\cite{Volonteri:2008gj} and will be
observable by
LISA,
while collision of  intermediate mass BHs and solar mass
BHs will lie in the sensitivity band of second and third generation
ground-based detectors~\cite{Mandel:2007hi, Mandel:2008bc, Will:2004fj}.

\vspace{.1in}
\noindent{\it Fully Nonlinear Numerical Simulations:}
In Table~\ref{table:ID}  we give the initial data parameters for our
$q=1/100$ BHB simulations. We evolved this BHB data set using 
the {\sc LazEv}~\cite{Zlochower:2005bj}
implementation of the moving puncture
approach~\cite{Campanelli:2005dd,Baker:2005vv}.  
Our code used the Cactus/Einstein toolkit~\cite{Cactusweb,
einsteintoolkit} and the 
Carpet~\cite{Schnetter-etal-03b} mesh refinement driver to
provide a `moving boxes' style mesh refinement. 
We use {\sc AHFinderDirect}~\cite{Thornburg2003:AH-finding} to locate
apparent horizons.  We measure the magnitude of the horizon spin using
the Isolated Horizon algorithm detailed in~\cite{Dreyer02a}. 
\begin{table}
\caption{Initial data parameters. The punctures are located on the
$x$-axis at positions $x_1$ and $x_2$, with puncture mass parameters
(not horizons masses) $m_1$ and $m_2$, and momentum $\pm\vec p$. 
The punctures have zero spin. The ADM mass $M_{\rm ADM}$ is $1M$ and $q=0.01000004$.}
\label{table:ID}
\begin{ruledtabular}
\begin{tabular}{ll|ll|ll}
$x_1$ & 4.95256  & $x_2$ & -0.0474374 &  $p_x$ & -0.0000102652 \\
$p_y$ & 0.00672262 &  $m_1$ & 0.00868947 & $m_2$  & 0.989619 \\
\end{tabular}
\end{ruledtabular}
\end{table}

We obtain accurate, convergent waveforms and horizon parameters by
evolving this system in conjunction with a modified 1+log lapse and a
modified Gamma-driver shift
condition~\cite{Alcubierre02a,Campanelli:2005dd}, and an initial lapse
$\alpha(t=0) = 2/(1+\psi_{BL}^{4})$.  The lapse and shift are evolved
with
$(\partial_t - \beta^i \partial_i) \alpha = - 2 \alpha K$, $\partial_t \beta^a = \frac34 \tilde \Gamma^a - \eta(x^k,t)\, \beta^a$,
where different functional dependences for $\eta(x^k,t)$ have been
proposed in 
\cite{Alcubierre:2004bm, Zlochower:2005bj, Mueller:2009jx, Mueller:2010bu, Schnetter:2010cz,Alic:2010wu}. 
Here we use a modification of the form proposed
in~\cite{Mueller:2009jx},
$\eta(x^k,t) =  R_0 \sqrt{\tilde\gamma^{ij}\partial_i W \partial_j W }/( \left(1 - W^a\right)^b)$,
where we chose $R_0=1.31$ and $W$ is the evolved conformal factor.
The above gauge condition is inspired by, but differs from 
Ref.~\cite{Mueller:2009jx} between the BHs and in the outer 
zones when $a\neq1$ and $b\neq2$.
Once the conformal factor settles down to its asymptotic
$\psi=C/\sqrt{r} + O(1)$ form near the puncture, $\eta$ will have the
form  $\eta = (R_0/C^2) ( 1+ b (r/C^2)^a)$ near the puncture and
$\eta= R_0 r^{b-2} M/(a M)^b$ as $r\to \infty$. Our exploration
of the $(a,b)$ parameters showed that the (1,2) case leads to numerical
instabilities on coarse grids, while the (2,1) and (1,1) cases 
lead to noisy waveforms and slower gauge speeds.
 In practice we used
$a=2$ and $b=2$, which reduces $\eta$ by a factor of $4$ at infinity
when compared to the original version of this gauge proposed
by~\cite{Mueller:2009jx}.
 We note that if we set $b=1$ then $\eta$
will have a $1/r$ falloff at $r=\infty$ as suggested
by~\cite{Schnetter:2010cz}.

In order to chose the width of the refinement levels closest to the small
BH, we examine the potentials for perturbations about a nonspinning
BH. The idea is that we need to model the curvature and
the gravitational radiation emitted by the small BH (which drives
the merger, and hence the physics). At the zeros of the
derivative of the potentials, the variations are minimized. Furthermore
the separations between zeros increases, naturally leading to a choice
of small-width, high resolution grids between the first zeros, one
step lower in resolution between the second two, followed by a
sequence of coarser grids.
According to Chandrasekhar \cite[p160]{Chandrasekhar83} the even/odd 
$(\pm)$ parity effective
potentials of a Schwarzschild BH can be written as
\begin{equation}
V^\pm_\ell=\pm6M\frac{df}{dr^*}+(6M)^2\,f^2+4\lambda(\lambda+1)\,f,
\end{equation}
where
\begin{equation}
f=\frac{(r-2M)}{2r^2(\lambda r+3M)},\quad\lambda=\frac12(\ell+2)(\ell-1).
\end{equation}
Note that both potentials are numerically very close to each other, hence we
consider the vanishing of the derivative of the average of the two (in isotropic coordinates $R$)
when constructing the grid.
For $\ell=2$ and $M=1$ this takes the explicit form
\begin{eqnarray}
&&\left.\frac{d(V^++V^-)}{dR}\right|_{\ell=2,M=1}=\\
&&-384\,{\frac {R \left( 2\,R-1 \right)  \left( 16\,{R}^{4}-4\,{R}^{3}- 
60\,{R}^{2}-R+1 \right)}{ \left( 2\,R+1 \right) ^{9
} \left( 4\,{R}^{2}+10\,R+1 \right) ^{3}}}\times\nonumber\\
&&\left( 64{R}^{6}+288{R}^{5}+480{R}^{4}+
256{R}^{3}+120{R}^{2}+18R+1 \right).\nonumber
\end{eqnarray}
Ideally, we would like to place the AMR boundaries around the small
BH near the zeros of
this function ($R/m=0.0,\ 0.1207998431,\ 0.5,\ 2.069539112$)
 The location of the zeros for $\ell>2$
changes little from the above figures. Since we do not want
to over-resolve the interior, this suggests that 
the first grid level should cover the whole small hole up to its horizon,
and the next grid level up to 4 times the initial horizon radius 
of the small hole in the initial quasi-isotropic coordinates.

In Ref.~\cite{Lousto:2010qx} we provide an alternative
method of extrapolation of waveforms based on a perturbative propagation
of the asymptotic form of $\psi_4$ at large distances,
leading to the following simple expression
\begin{eqnarray}
&&\lim_{r\to\infty}[r \,\psi_{4}^{\ell m}(r,t)] = \\
&&\left[r \,\psi_{4}^{\ell m}(r,t) 
- \lambda \int_0^t dt \, \psi_{4}^{\ell m}(r,t)
\right]_{r=r_{\rm Obs}} 
+ O(r_{\rm Obs}^{-2}),\nonumber
\label{eq:asymtpsi4ext}
\end{eqnarray}
where $r_{\rm Obs}$ is the approximate areal radius of the
sphere. 
This formula is applicable for $r_{\rm Obs}\gtrsim100M$. And
note that it is also important to remove the low frequency 
components \cite{Reisswig:2010di} in $\psi_4$ (since it is inside an integral).

\vspace{.1in}
\noindent{\it Results and Analysis:}
 Our simulation used
15 levels of refinement (around the smaller components), with central
 resolutions as high as $M/7078$, and 9 levels of refinement around
the larger component. The outer boundaries were located at $400M$ and
the resolution in the boundary zone was $h=2.3148M$ for our finest
resolution run. The BHB performs $\sim2$ orbits prior to merger 
[as seen by the formation of a common apparent horizon (CAH)], which
occurs roughly $160M$ after the start of the simulation. In terms of
computational expense, a medium resolution run requires 500,000 SU
and approximately one month of runtime. In order to reduce the
total runtime, we used an aggressive choice of CFL ($dt=1/2 h$),
which leads to significant BH mass loss when compared to
$dt=1/4 h$.

\begin{figure}
  \includegraphics[width=2.2in,height=2.5in]{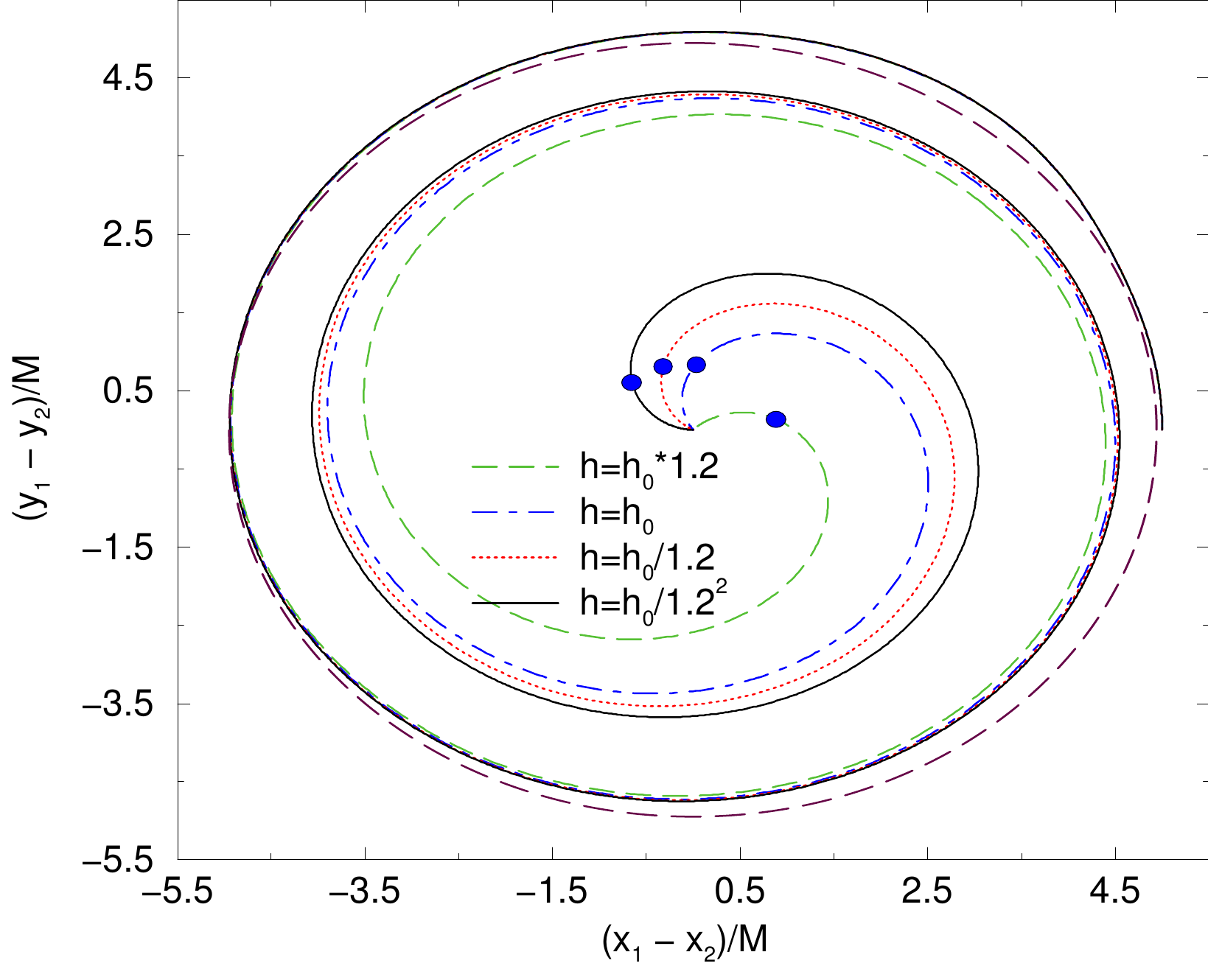}
  \caption{An $xy$ projection of the trajectories for the two
    highest resolutions of the $q=1/100$ configuration. The 
  dotted circle corresponds to the ISCO radius while the small filled-in circle
 corresponds to the point on the trajectory where a common horizon is first
detected. Note the initial ``jump'' in radius (see
Fig.~\ref{fig:rtrack}) due to the initial data radiation content.}
  \label{fig:track}
\end{figure}

\begin{figure}
  \includegraphics[width=2.2in]{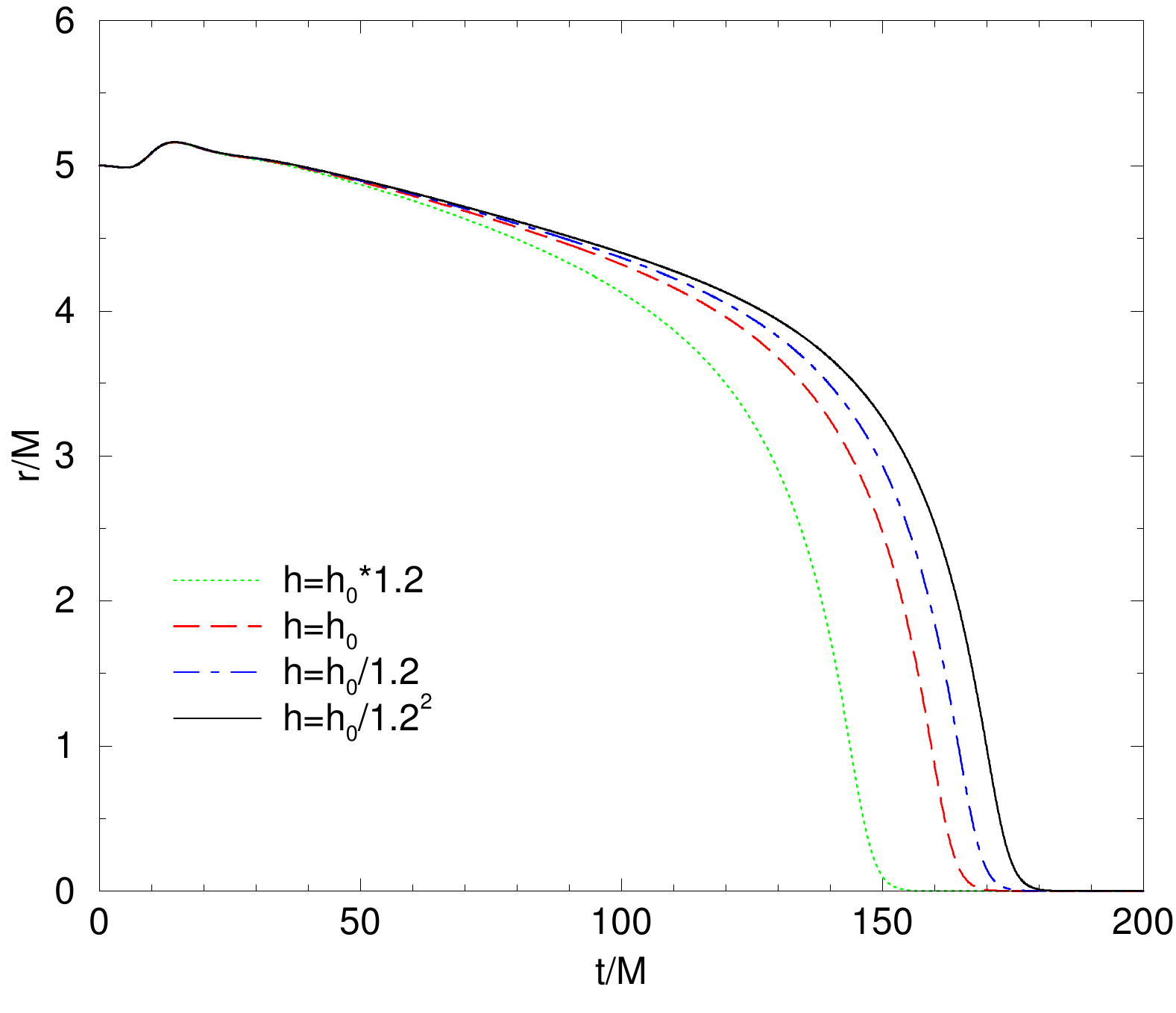}
  \caption{The orbital radius as a function of time and resolution for
the $q=1/100$ configuration. Note
    the initial ``jump'' in the orbit due to the initial data.}
  \label{fig:rtrack}
\end{figure}

\begin{figure}
  \includegraphics[width=2.6in]{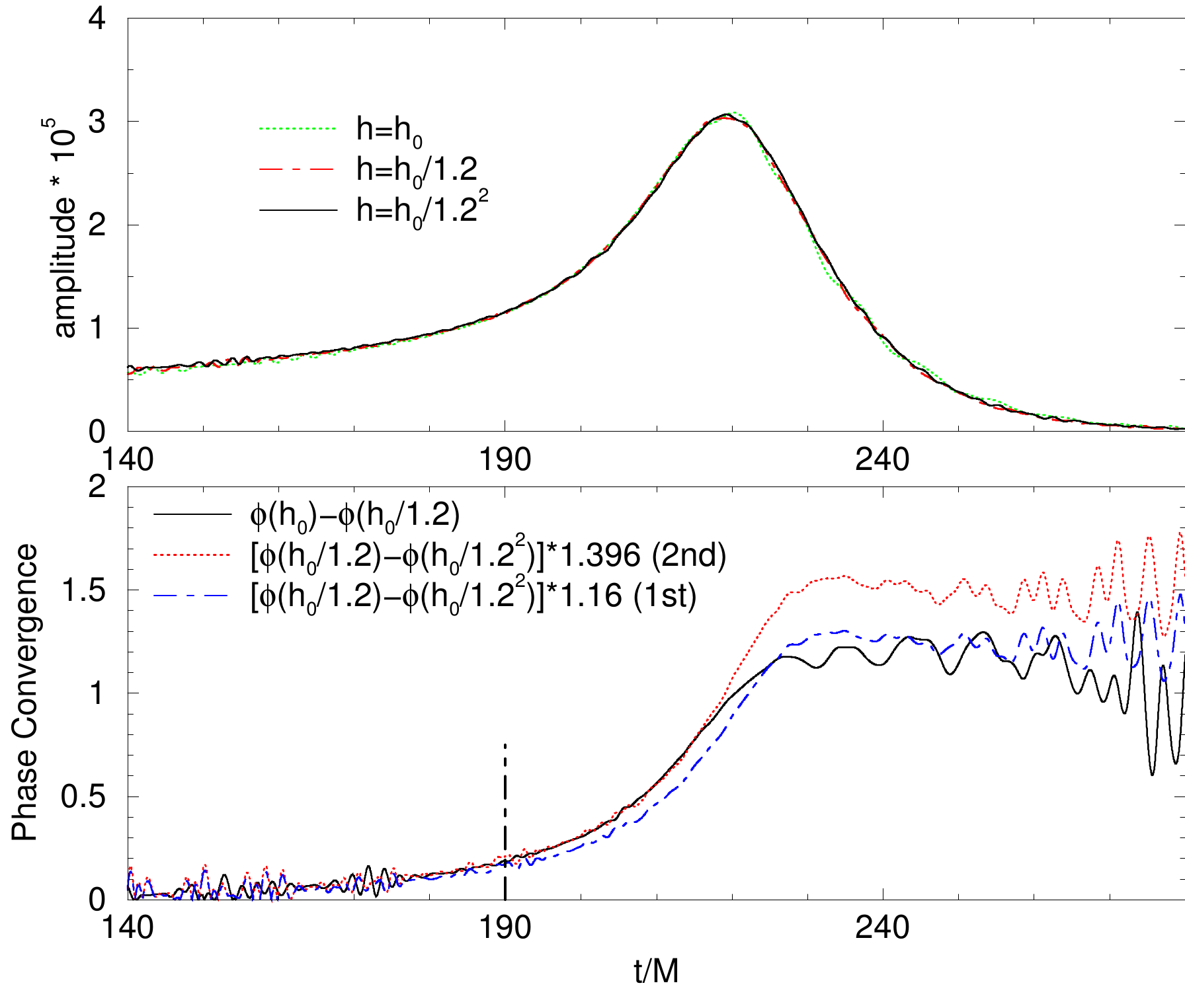}
  \caption{Convergence of the amplitude and phase of the $(\ell=2,m=2)$ mode
    of $\psi_4$. The phase converges to second order prior to the peak
    in the amplitude. The vertical line shows the point when $\omega=0.2$.
    Note the good agreement in amplitude (the curves have been translated).
    The phase error at $\omega=0.2$ is $0.44$ radians.
}
  \label{fig:wave}
\end{figure}

Table~\ref{table:Remnant} shows the results of evolution. We note that
the smaller BH mass is conserved to within $0.23\%$
during the inspiral and plunge phases,
 while the mass of the larger BH is conserved
to within $0.003\%$. In Fig.~\ref{fig:track} we show the $xy$
projection of the orbital trajectories for the two highest resolution
runs. From the figure we can see that the initial jump in the orbit
pushes the binary slightly outside the ISCO, leading to an additional
orbit. In
 Fig.~\ref{fig:rtrack} we show the orbital radius as a function of time
and resolution. Note that the orbital radius superconverges at low
resolution and converges quadratically at high resolution ( this quadratic
error may be due to time prolongation effects, as well 
as effects due to an aggressive choice of CFL factor). In Fig.~\ref{fig:wave}
we show amplitude, as well as phase convergence, of the $(\ell=2,m=2)$
 mode of $\psi_4$. The reduced order of convergence is due to an aggressive
choice of CFL factor.

The apparent superconvergence in the trajectories and waveforms when
considering the three coarsest resolutions is indicative that the
lowest resolution is just entering the convergence regime. That is,
this resolution cannot be far from the convergence regime because all
four resolutions lie in a monotonic convergence sequence. And
importantly, the deviations between the next three resolutions are very
small compared to the deviation between the lowest two resolutions,
indicating that these three resolutions are safely inside the
convergence regime.

\begin{table}
\caption{Remnant horizon parameters and radiated energy-momentum. Here
we provide $\delta M_{H}^* = M_{\rm ADM} - M_H$ and $\delta S_{H}^* =
J_{ADM} - S_H$, which are small numbers obtained by
taking the difference between two much larger numbers. The
calculation of $\delta S_{H}^*$ is relatively inaccurate because it
requires an extrapolation to infinite
resolution.}
\label{table:Remnant}
\begin{ruledtabular}
\begin{tabular}{ll|ll|ll}
$10^{5} E_{rad}$ & $6.0 \pm 0.1$ & $10^{5}\delta M_{H}^*$ & $7.0 \pm 1.0$ & $100 \alpha$ & $3.33 \pm 0.02$ \\
 $10^{4} J_{rad}$ & $5.0 \pm 0.2$  & $10^{4}\delta S_{H}^*$  & $3.0 \pm 2.0$
 & $V_{\rm kick}$ & $1.07\pm0.05 \KMS$\\
\end{tabular}
\end{ruledtabular}
\end{table}

\begin{figure}
  \includegraphics[width=2.4in]{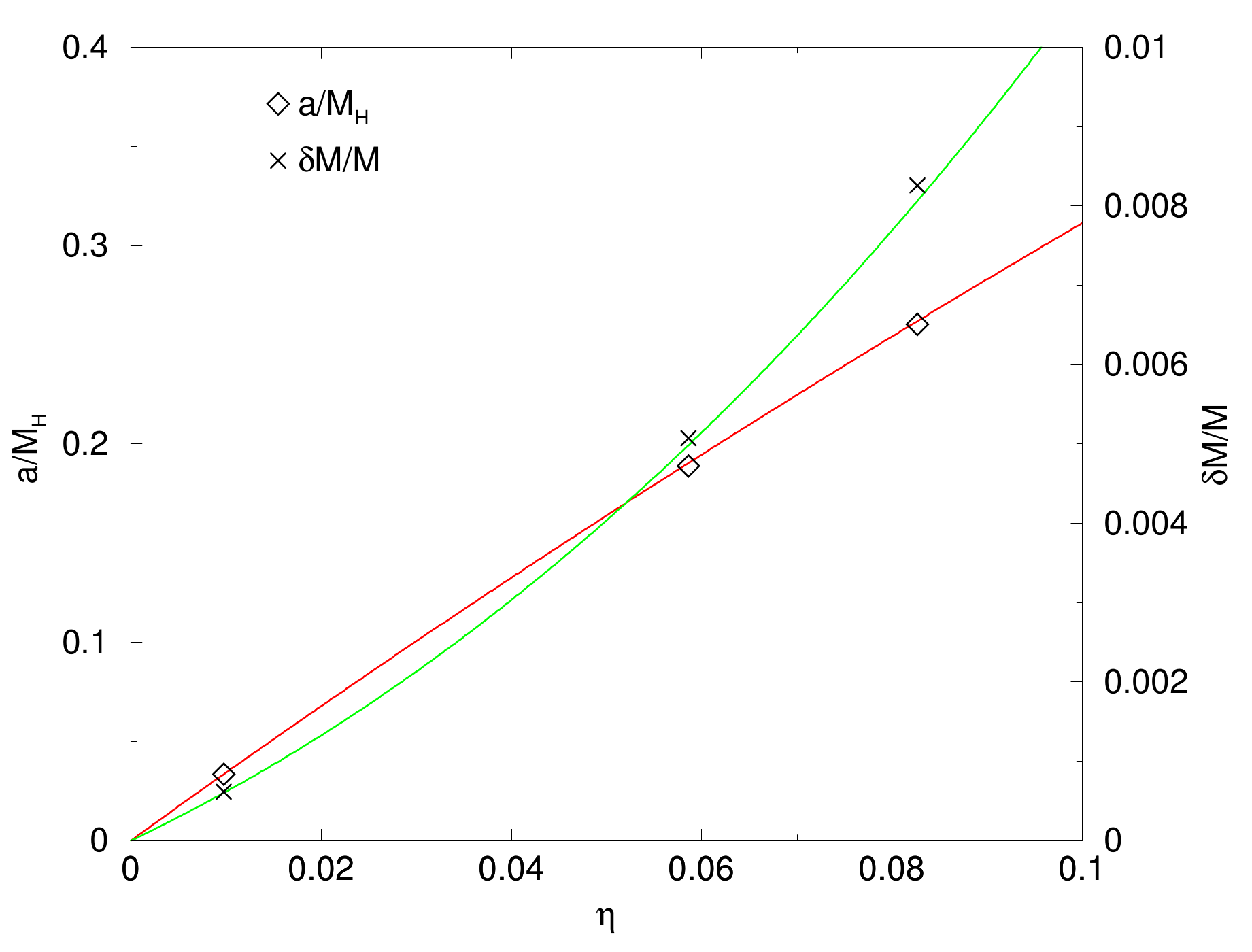}
  \caption{The remnant spin $a/M_H$ and the radiated mass from infinite
separation $\delta M/M$, as a function of the symmetric mass ratio
$\eta = q/(1+q)^2$,
for $q=1/10$, $1/15$, $1/100$, as well as the empirical formula
prediction.}
  \label{fig:remnant}
\end{figure}

Finally, in Fig.~\ref{fig:remnant} and Table~\ref{table:empir} 
we show the remnant spins and total
radiated mass as a function of mass ratio \cite{Lousto:2010qx}
for $q=1/10, 1/15, 1/100$
and the predictions based on our empirical formula~\cite{Lousto:2009mf}.
Note that no fitting is involved in this figure.

\begin{table}
\caption{Remnant spin and total radiated mass (starting from infinite
separation) as a function of mass ratio $q$ as measured in our
simulations and as predicted by our empirical formulae~\cite{Lousto:2009mf}.}
\label{table:empir}
\begin{ruledtabular}
\begin{tabular}{llll}
$q$ & 1/10 & 1/15 & 1/100 \\
\hline
$\alpha$ (Computed) & 0.2603 & 0.18875 & 0.0333 \\
$\alpha$ (Predicted) & 0.2618 & 0.1903 & 0.03358 \\
$\delta M$ (Computed)  & 0.00826 & 0.00507 & 0.000618 \\
$\delta M$ (Predicted) & 0.00806 & 0.00498 & 0.000604
\end{tabular}
\end{ruledtabular}
\end{table}

The amount of energy and angular momentum radiated when
the $(2,2)$ mode frequency is larger than $M\omega_{2,2} > 0.167$
is given by (adding up to $\ell=4$ modes)
$\delta E/M = 0.000047\pm0.000001$ and $\delta J/M^2 =
0.00034\pm0.00001$, which agrees to within $4\%$ with the particle
limit predictions of $\delta E/M = 0.47\eta^2$ and $\delta J/M^2 =
3.44\eta^2$~\cite{Bernuzzi:2010ty}.

\vspace{.1in}
\noindent{\it Conclusions and Discussion:}
We have successfully evolved a 1:100 BHB system for the last two
orbits before merger and down to the final Kerr hole remnant. 
We have achieved this within the moving punctures approach
by adapting the gamma-driver shift condition with a variable damping term.
Also crucial for evolutions is an optimal choice of the mesh refinement 
structure around
the small BH. We used the Regge-Wheeler-Zerilli potentials to guide
the setting up of the initial grids. This helps optimizing the large resources
required to evolve small $q$ binaries. 
The numerical convergence of the waveforms displayed here,
and the successful comparisons with perturbative results 
\cite{Lousto:2010tb, Lousto:2010qx},
show this approach is validated in the intermediate mass ratio regime,
and can be applied to even smaller $q$'s 
(and larger initial separations into the post-Newtonian regime).

The feasibility of simulating extreme mass ratios by purely 
fully nonlinear numerical
methods, as demonstrated in this work,
allows us to look more optimistically at the task of generating a bank of
templates for second and third generation ground-based detectors
and LISA. 
Methods like those described in \cite{Lousto:2010tb,Lousto:2010qx},
that combine NR and perturbative techniques can be
used to speed up the generation of those templates. And finally
we now also
have a direct way of validating self-force computations \cite{Barack:2010tm}.

The techniques presented here would appear to apply in a 
straightforward manner to 
even smaller mass ratios $q$ and to initially spinning BHs.
Fine tuning of the quasicircular orbital parameters plays an important
role in preparing these runs, given the very low level of gravitational
radiation they generate. So far we see that the method \cite{Pfeiffer:2007yz}
developed for equal mass BHBs to lower the eccentricity seems to work, but
it requires extra runs for initial experimentation.
Hence it would be important to evolve initial data with lower spurious 
radiation content and some true inspiral wave information \cite{Kelly:2009js}.

\acknowledgments
We gratefully acknowledge the NSF for financial support from Grants
No. PHY-0722315, No. PHY-0653303, No. PHY-0714388, No. PHY-0722703,
No. DMS-0820923, No. PHY-0929114, No. PHY-0969855, No. PHY-0903782,
No. CDI-1028087; and NASA for financial support from NASA Grants 
No. 07-ATFP07-0158 and No. HST-AR-11763.  Computational resources were
provided by the Ranger cluster at TACC (Teragrid allocation TG-PHY060027N) 
and by NewHorizons at RIT.

\bibliographystyle{apsrev}
\bibliography{../../../numrelcvs/Bibtex/references}

\begin{thebibliography}{35}
\expandafter\ifx\csname natexlab\endcsname\relax\def\natexlab#1{#1}\fi
\expandafter\ifx\csname bibnamefont\endcsname\relax
  \def\bibnamefont#1{#1}\fi
\expandafter\ifx\csname bibfnamefont\endcsname\relax
  \def\bibfnamefont#1{#1}\fi
\expandafter\ifx\csname citenamefont\endcsname\relax
  \def\citenamefont#1{#1}\fi
\expandafter\ifx\csname url\endcsname\relax
  \def\url#1{\texttt{#1}}\fi
\expandafter\ifx\csname urlprefix\endcsname\relax\def\urlprefix{URL }\fi
\providecommand{\bibinfo}[2]{#2}
\providecommand{\eprint}[2][]{\url{#2}}

\bibitem[{\citenamefont{Regge and Wheeler}(1957)}]{Regge57}
\bibinfo{author}{\bibfnamefont{T.}~\bibnamefont{Regge}} \bibnamefont{and}
  \bibinfo{author}{\bibfnamefont{J.}~\bibnamefont{Wheeler}},
  \bibinfo{journal}{Phys. Rev.} \textbf{\bibinfo{volume}{108}},
  \bibinfo{pages}{1063} (\bibinfo{year}{1957}).

\bibitem[{\citenamefont{Zerilli}(1970)}]{Zerilli:1971wd}
\bibinfo{author}{\bibfnamefont{F.~J.} \bibnamefont{Zerilli}},
  \bibinfo{journal}{Phys. Rev. D} \textbf{\bibinfo{volume}{2}},
  \bibinfo{pages}{2141} (\bibinfo{year}{1970}).

\bibitem[{\citenamefont{Teukolsky}(1973)}]{Teukolsky:1973ha}
\bibinfo{author}{\bibfnamefont{S.~A.} \bibnamefont{Teukolsky}},
  \bibinfo{journal}{Astrophys. J.} \textbf{\bibinfo{volume}{185}},
  \bibinfo{pages}{635} (\bibinfo{year}{1973}).

\bibitem[{\citenamefont{Mino et~al.}(1997)\citenamefont{Mino, Sasaki, and
  Tanaka}}]{Mino:1996nk}
\bibinfo{author}{\bibfnamefont{Y.}~\bibnamefont{Mino}},
  \bibinfo{author}{\bibfnamefont{M.}~\bibnamefont{Sasaki}}, \bibnamefont{and}
  \bibinfo{author}{\bibfnamefont{T.}~\bibnamefont{Tanaka}},
  \bibinfo{journal}{Phys. Rev.} \textbf{\bibinfo{volume}{D55}},
  \bibinfo{pages}{3457} (\bibinfo{year}{1997}).

\bibitem[{\citenamefont{Quinn and Wald}(1997)}]{Quinn:1996am}
\bibinfo{author}{\bibfnamefont{T.~C.} \bibnamefont{Quinn}} \bibnamefont{and}
  \bibinfo{author}{\bibfnamefont{R.~M.} \bibnamefont{Wald}},
  \bibinfo{journal}{Phys. Rev.} \textbf{\bibinfo{volume}{D56}},
  \bibinfo{pages}{3381} (\bibinfo{year}{1997}).

\bibitem[{\citenamefont{Barack and Sago}(2010)}]{Barack:2010tm}
\bibinfo{author}{\bibfnamefont{L.}~\bibnamefont{Barack}} \bibnamefont{and}
  \bibinfo{author}{\bibfnamefont{N.}~\bibnamefont{Sago}},
  \bibinfo{journal}{Phys. Rev.} \textbf{\bibinfo{volume}{D81}},
  \bibinfo{pages}{084021} (\bibinfo{year}{2010}).

\bibitem[{\citenamefont{Pretorius}(2005)}]{Pretorius:2005gq}
\bibinfo{author}{\bibfnamefont{F.}~\bibnamefont{Pretorius}},
  \bibinfo{journal}{Phys. Rev. Lett.} \textbf{\bibinfo{volume}{95}},
  \bibinfo{pages}{121101} (\bibinfo{year}{2005}).

\bibitem[{\citenamefont{Campanelli et~al.}(2006)\citenamefont{Campanelli,
  Lousto, Marronetti, and Zlochower}}]{Campanelli:2005dd}
\bibinfo{author}{\bibfnamefont{M.}~\bibnamefont{Campanelli}},
  \bibinfo{author}{\bibfnamefont{C.~O.} \bibnamefont{Lousto}},
  \bibinfo{author}{\bibfnamefont{P.}~\bibnamefont{Marronetti}},
  \bibnamefont{and}
  \bibinfo{author}{\bibfnamefont{Y.}~\bibnamefont{Zlochower}},
  \bibinfo{journal}{Phys. Rev. Lett.} \textbf{\bibinfo{volume}{96}},
  \bibinfo{pages}{111101} (\bibinfo{year}{2006}).

\bibitem[{\citenamefont{Baker et~al.}(2006)\citenamefont{Baker, Centrella,
  Choi, Koppitz, and van Meter}}]{Baker:2005vv}
\bibinfo{author}{\bibfnamefont{J.~G.} \bibnamefont{Baker}},
  \bibinfo{author}{\bibfnamefont{J.}~\bibnamefont{Centrella}},
  \bibinfo{author}{\bibfnamefont{D.-I.} \bibnamefont{Choi}},
  \bibinfo{author}{\bibfnamefont{M.}~\bibnamefont{Koppitz}}, \bibnamefont{and}
  \bibinfo{author}{\bibfnamefont{J.}~\bibnamefont{van Meter}},
  \bibinfo{journal}{Phys. Rev. Lett.} \textbf{\bibinfo{volume}{96}},
  \bibinfo{pages}{111102} (\bibinfo{year}{2006}).

\bibitem[{\citenamefont{Lousto and Zlochower}(2009)}]{Lousto:2008dn}
\bibinfo{author}{\bibfnamefont{C.~O.} \bibnamefont{Lousto}} \bibnamefont{and}
  \bibinfo{author}{\bibfnamefont{Y.}~\bibnamefont{Zlochower}},
  \bibinfo{journal}{Phys. Rev. D} \textbf{\bibinfo{volume}{79}},
  \bibinfo{pages}{064018} (\bibinfo{year}{2009}).

\bibitem[{\citenamefont{Gonzalez et~al.}(2009)\citenamefont{Gonzalez, Sperhake,
  and Brugmann}}]{Gonzalez:2008bi}
\bibinfo{author}{\bibfnamefont{J.~A.} \bibnamefont{Gonzalez}},
  \bibinfo{author}{\bibfnamefont{U.}~\bibnamefont{Sperhake}}, \bibnamefont{and}
  \bibinfo{author}{\bibfnamefont{B.}~\bibnamefont{Brugmann}},
  \bibinfo{journal}{Phys. Rev.} \textbf{\bibinfo{volume}{D79}},
  \bibinfo{pages}{124006} (\bibinfo{year}{2009}).

\bibitem[{\citenamefont{Lousto et~al.}(2010{\natexlab{a}})\citenamefont{Lousto,
  Nakano, Zlochower, and Campanelli}}]{Lousto:2010tb}
\bibinfo{author}{\bibfnamefont{C.~O.} \bibnamefont{Lousto}},
  \bibinfo{author}{\bibfnamefont{H.}~\bibnamefont{Nakano}},
  \bibinfo{author}{\bibfnamefont{Y.}~\bibnamefont{Zlochower}},
  \bibnamefont{and}
  \bibinfo{author}{\bibfnamefont{M.}~\bibnamefont{Campanelli}},
  \bibinfo{journal}{Phys. Rev. Lett.} \textbf{\bibinfo{volume}{104}},
  \bibinfo{pages}{211101} (\bibinfo{year}{2010}{\natexlab{a}}).

\bibitem[{\citenamefont{Lousto et~al.}(2010{\natexlab{b}})\citenamefont{Lousto,
  Nakano, Zlochower, and Campanelli}}]{Lousto:2010qx}
\bibinfo{author}{\bibfnamefont{C.~O.} \bibnamefont{Lousto}},
  \bibinfo{author}{\bibfnamefont{H.}~\bibnamefont{Nakano}},
  \bibinfo{author}{\bibfnamefont{Y.}~\bibnamefont{Zlochower}},
  \bibnamefont{and}
  \bibinfo{author}{\bibfnamefont{M.}~\bibnamefont{Campanelli}}
  (\bibinfo{year}{2010}{\natexlab{b}}), \eprint{1008.4360}.

\bibitem[{\citenamefont{{Volonteri} and {Madau}}(2008)}]{Volonteri:2008gj}
\bibinfo{author}{\bibfnamefont{M.}~\bibnamefont{{Volonteri}}} \bibnamefont{and}
  \bibinfo{author}{\bibfnamefont{P.}~\bibnamefont{{Madau}}},
  \bibinfo{journal}{Astrophys. J.} \textbf{\bibinfo{volume}{687}},
  \bibinfo{pages}{L57} (\bibinfo{year}{2008}).

\bibitem[{\citenamefont{Mandel et~al.}(2008)\citenamefont{Mandel, Brown, Gair,
  and Miller}}]{Mandel:2007hi}
\bibinfo{author}{\bibfnamefont{I.}~\bibnamefont{Mandel}},
  \bibinfo{author}{\bibfnamefont{D.~A.} \bibnamefont{Brown}},
  \bibinfo{author}{\bibfnamefont{J.~R.} \bibnamefont{Gair}}, \bibnamefont{and}
  \bibinfo{author}{\bibfnamefont{M.~C.} \bibnamefont{Miller}},
  \bibinfo{journal}{Astrophys. J.} \textbf{\bibinfo{volume}{681}},
  \bibinfo{pages}{1431} (\bibinfo{year}{2008}).

\bibitem[{\citenamefont{Mandel and Gair}(2009)}]{Mandel:2008bc}
\bibinfo{author}{\bibfnamefont{I.}~\bibnamefont{Mandel}} \bibnamefont{and}
  \bibinfo{author}{\bibfnamefont{J.~R.} \bibnamefont{Gair}},
  \bibinfo{journal}{Class. Quant. Grav.} \textbf{\bibinfo{volume}{26}},
  \bibinfo{pages}{094036} (\bibinfo{year}{2009}).

\bibitem[{\citenamefont{Will}(2004)}]{Will:2004fj}
\bibinfo{author}{\bibfnamefont{C.~M.} \bibnamefont{Will}},
  \bibinfo{journal}{Astrophys. J.} \textbf{\bibinfo{volume}{611}},
  \bibinfo{pages}{1080} (\bibinfo{year}{2004}).

\bibitem[{\citenamefont{Zlochower et~al.}(2005)\citenamefont{Zlochower, Baker,
  Campanelli, and Lousto}}]{Zlochower:2005bj}
\bibinfo{author}{\bibfnamefont{Y.}~\bibnamefont{Zlochower}},
  \bibinfo{author}{\bibfnamefont{J.~G.} \bibnamefont{Baker}},
  \bibinfo{author}{\bibfnamefont{M.}~\bibnamefont{Campanelli}},
  \bibnamefont{and} \bibinfo{author}{\bibfnamefont{C.~O.}
  \bibnamefont{Lousto}}, \bibinfo{journal}{Phys. Rev. D}
  \textbf{\bibinfo{volume}{72}}, \bibinfo{pages}{024021}
  (\bibinfo{year}{2005}).

\bibitem[{Cac()}]{Cactusweb}
\emph{\bibinfo{title}{Cactus: {\tt http://www.cactuscode.org}}}.

\bibitem[{ein()}]{einsteintoolkit}
\bibinfo{note}{Einstein Toolkit home page: {\tt http://einsteintoolkit.org}}.

\bibitem[{\citenamefont{Schnetter et~al.}(2004)\citenamefont{Schnetter, Hawley,
  and Hawke}}]{Schnetter-etal-03b}
\bibinfo{author}{\bibfnamefont{E.}~\bibnamefont{Schnetter}},
  \bibinfo{author}{\bibfnamefont{S.~H.} \bibnamefont{Hawley}},
  \bibnamefont{and} \bibinfo{author}{\bibfnamefont{I.}~\bibnamefont{Hawke}},
  \bibinfo{journal}{Class. Quantum Grav.} \textbf{\bibinfo{volume}{21}},
  \bibinfo{pages}{1465} (\bibinfo{year}{2004}).

\bibitem[{\citenamefont{Thornburg}(2004)}]{Thornburg2003:AH-finding}
\bibinfo{author}{\bibfnamefont{J.}~\bibnamefont{Thornburg}},
  \bibinfo{journal}{Class. Quantum Grav.} \textbf{\bibinfo{volume}{21}},
  \bibinfo{pages}{743} (\bibinfo{year}{2004}).

\bibitem[{\citenamefont{Dreyer et~al.}(2003)\citenamefont{Dreyer, Krishnan,
  Shoemaker, and Schnetter}}]{Dreyer02a}
\bibinfo{author}{\bibfnamefont{O.}~\bibnamefont{Dreyer}},
  \bibinfo{author}{\bibfnamefont{B.}~\bibnamefont{Krishnan}},
  \bibinfo{author}{\bibfnamefont{D.}~\bibnamefont{Shoemaker}},
  \bibnamefont{and}
  \bibinfo{author}{\bibfnamefont{E.}~\bibnamefont{Schnetter}},
  \bibinfo{journal}{Phys. Rev. D} \textbf{\bibinfo{volume}{67}},
  \bibinfo{pages}{024018} (\bibinfo{year}{2003}).

\bibitem[{\citenamefont{Alcubierre et~al.}(2003)\citenamefont{Alcubierre,
  Br\"ugmann, Diener, Koppitz, Pollney, Seidel, and Takahashi}}]{Alcubierre02a}
\bibinfo{author}{\bibfnamefont{M.}~\bibnamefont{Alcubierre}},
  \bibinfo{author}{\bibfnamefont{B.}~\bibnamefont{Br\"ugmann}},
  \bibinfo{author}{\bibfnamefont{P.}~\bibnamefont{Diener}},
  \bibinfo{author}{\bibfnamefont{M.}~\bibnamefont{Koppitz}},
  \bibinfo{author}{\bibfnamefont{D.}~\bibnamefont{Pollney}},
  \bibinfo{author}{\bibfnamefont{E.}~\bibnamefont{Seidel}}, \bibnamefont{and}
  \bibinfo{author}{\bibfnamefont{R.}~\bibnamefont{Takahashi}},
  \bibinfo{journal}{Phys. Rev. D} \textbf{\bibinfo{volume}{67}},
  \bibinfo{pages}{084023} (\bibinfo{year}{2003}).

\bibitem[{\citenamefont{Alcubierre et~al.}(2004)}]{Alcubierre:2004bm}
\bibinfo{author}{\bibfnamefont{M.}~\bibnamefont{Alcubierre}}
  \bibnamefont{et~al.} (\bibinfo{year}{2004}), \eprint{gr-qc/0411137}.

\bibitem[{\citenamefont{Mueller and Bruegmann}(2010)}]{Mueller:2009jx}
\bibinfo{author}{\bibfnamefont{D.}~\bibnamefont{Mueller}} \bibnamefont{and}
  \bibinfo{author}{\bibfnamefont{B.}~\bibnamefont{Bruegmann}},
  \bibinfo{journal}{Class. Quant. Grav.} \textbf{\bibinfo{volume}{27}},
  \bibinfo{pages}{114008} (\bibinfo{year}{2010}).

\bibitem[{\citenamefont{Mueller et~al.}(2010)\citenamefont{Mueller, Grigsby,
  and Bruegmann}}]{Mueller:2010bu}
\bibinfo{author}{\bibfnamefont{D.}~\bibnamefont{Mueller}},
  \bibinfo{author}{\bibfnamefont{J.}~\bibnamefont{Grigsby}}, \bibnamefont{and}
  \bibinfo{author}{\bibfnamefont{B.}~\bibnamefont{Bruegmann}},
  \bibinfo{journal}{Phys. Rev.} \textbf{\bibinfo{volume}{D82}},
  \bibinfo{pages}{064004} (\bibinfo{year}{2010}).

\bibitem[{\citenamefont{Schnetter}(2010)}]{Schnetter:2010cz}
\bibinfo{author}{\bibfnamefont{E.}~\bibnamefont{Schnetter}},
  \bibinfo{journal}{Class. Quant. Grav.} \textbf{\bibinfo{volume}{27}},
  \bibinfo{pages}{167001} (\bibinfo{year}{2010}).

\bibitem[{\citenamefont{Alic et~al.}(2010)\citenamefont{Alic, Rezzolla, Hinder,
  and Mosta}}]{Alic:2010wu}
\bibinfo{author}{\bibfnamefont{D.}~\bibnamefont{Alic}},
  \bibinfo{author}{\bibfnamefont{L.}~\bibnamefont{Rezzolla}},
  \bibinfo{author}{\bibfnamefont{I.}~\bibnamefont{Hinder}}, \bibnamefont{and}
  \bibinfo{author}{\bibfnamefont{P.}~\bibnamefont{Mosta}}
  (\bibinfo{year}{2010}), \eprint{1008.2212}.

\bibitem[{\citenamefont{Chandrasekhar}(1983)}]{Chandrasekhar83}
\bibinfo{author}{\bibfnamefont{S.}~\bibnamefont{Chandrasekhar}},
  \emph{\bibinfo{title}{The Mathematical Theory of Black Holes}}
  (\bibinfo{publisher}{Oxford University Press}, \bibinfo{address}{Oxford,
  U.K.}, \bibinfo{year}{1983}).

\bibitem[{\citenamefont{Reisswig and Pollney}(2010)}]{Reisswig:2010di}
\bibinfo{author}{\bibfnamefont{C.}~\bibnamefont{Reisswig}} \bibnamefont{and}
  \bibinfo{author}{\bibfnamefont{D.}~\bibnamefont{Pollney}}
  (\bibinfo{year}{2010}), \eprint{1006.1632}.

\bibitem[{\citenamefont{Lousto et~al.}(2010{\natexlab{c}})\citenamefont{Lousto,
  Campanelli, Zlochower, and Nakano}}]{Lousto:2009mf}
\bibinfo{author}{\bibfnamefont{C.~O.} \bibnamefont{Lousto}},
  \bibinfo{author}{\bibfnamefont{M.}~\bibnamefont{Campanelli}},
  \bibinfo{author}{\bibfnamefont{Y.}~\bibnamefont{Zlochower}},
  \bibnamefont{and} \bibinfo{author}{\bibfnamefont{H.}~\bibnamefont{Nakano}},
  \bibinfo{journal}{Class. Quant. Grav.} \textbf{\bibinfo{volume}{27}},
  \bibinfo{pages}{114006} (\bibinfo{year}{2010}{\natexlab{c}}).

\bibitem[{\citenamefont{Bernuzzi and Nagar}(2010)}]{Bernuzzi:2010ty}
\bibinfo{author}{\bibfnamefont{S.}~\bibnamefont{Bernuzzi}} \bibnamefont{and}
  \bibinfo{author}{\bibfnamefont{A.}~\bibnamefont{Nagar}},
  \bibinfo{journal}{Phys. Rev.} \textbf{\bibinfo{volume}{D81}},
  \bibinfo{pages}{084056} (\bibinfo{year}{2010}).

\bibitem[{\citenamefont{Pfeiffer et~al.}(2007)}]{Pfeiffer:2007yz}
\bibinfo{author}{\bibfnamefont{H.~P.} \bibnamefont{Pfeiffer}}
  \bibnamefont{et~al.}, \bibinfo{journal}{Class. Quant. Grav.}
  \textbf{\bibinfo{volume}{24}}, \bibinfo{pages}{S59} (\bibinfo{year}{2007}).

\bibitem[{\citenamefont{Kelly et~al.}(2010)\citenamefont{Kelly, Tichy,
  Zlochower, Campanelli, and Whiting}}]{Kelly:2009js}
\bibinfo{author}{\bibfnamefont{B.~J.} \bibnamefont{Kelly}},
  \bibinfo{author}{\bibfnamefont{W.}~\bibnamefont{Tichy}},
  \bibinfo{author}{\bibfnamefont{Y.}~\bibnamefont{Zlochower}},
  \bibinfo{author}{\bibfnamefont{M.}~\bibnamefont{Campanelli}},
  \bibnamefont{and} \bibinfo{author}{\bibfnamefont{B.~F.}
  \bibnamefont{Whiting}}, \bibinfo{journal}{Class. Quant. Grav.}
  \textbf{\bibinfo{volume}{27}}, \bibinfo{pages}{114005}
  (\bibinfo{year}{2010}).

\end{thebibliography}

\end{document}